# Using thermal boundary conditions to engineer the quantum state of a bulk magnet

*M.A. Schmidt[1], D.M. Silevitch[1], G. Aeppli[2], T.F. Rosenbaum[1,*]*

[1]The James Franck Institute and Department of Physics, The University of Chicago, Chicago, Illinois 60637, USA

[2]London Centre for Nanotechnology and Department of Physics and Astronomy, UCL, London, WC1E 6BT, UK

* Correspondence to:

   T.F. Rosenbaum
   The University of Chicago
   929 E. 57th Street
   Chicago, IL 60637
   773-702-7256
   tfr@uchicago.edu

**Abstract**

The degree of contact between a system and the external environment can alter dramatically its proclivity to quantum mechanical modes of relaxation. We show that controlling the thermal coupling of cubic centimeter-sized crystals of the Ising magnet LiHo$_x$Y$_{1-x}$F$_4$ to a heat bath can be used to tune the system between a glassy state dominated by thermal excitations over energy barriers and a state with the hallmarks of a quantum spin liquid. Application of a magnetic field transverse to the Ising axis introduces both random magnetic fields and quantum fluctuations, which can retard and speed the annealing process, respectively, thereby providing a mechanism for continuous tuning between the destination states. The non-linear response of the system explicitly demonstrates quantum interference between internal and external relaxation pathways.

**Significance Statement**

The interactions of a material with its environment can determine its behavior and induce changes of state. We show that at temperatures near absolute zero a magnetic material can be made more quantum mechanical by isolating it from the environment. Local clusters of spins within the material stay liquid and refuse to freeze. An oscillating magnetic field serves as an effective tool to address and manipulate these "protected" spin clusters, while a dc magnetic field can enhance the spin tunneling rate and lead to quantum speed-up. When the material is more strongly connected to a heat bath, the local magnetic clusters behave more classically and freeze en masse into a glassy state.

**Introduction**

The coupling of a sample to its environment is both a fundamental theoretical concept and a powerful experimental tool in classical thermodynamics. For quantum systems, contact between the internal degrees of freedom and the external world, often referred to as the "bath," can change the measured outcome completely. Typically, such experiments involve a small number of particles sensitive to subtle changes in the external incoherent environment, such as ultracold atoms confined in precisely controlled optical potentials (1-3). With the search for viable solid-state qubits for quantum computing, the control of bath-induced decoherence in solids also has become an important topic for engineers and condensed matter physicists. Approaches have centered on the nuclear spin bath (4-6), modifying it either with isotopic substitution (7) or radio frequency pulses (8), and on electrical control of the exchange interaction between electron spins in coupled quantum dots (9). The question of the importance of coupling to an external bath, as provided by a cryostat, has not been researched as intensively. Here, we show that by engineering the thermal boundary conditions for a macroscopic magnetic crystal, it is possible to select distinct low temperature states. Conditions of constant energy, as opposed to constant temperature, yield relatively fewer low energy contributions to the fluctuation spectrum and decouple the spin excitations responsible for that spectrum into separate oscillators. The experiments show the importance of thermal heat sinking for quantum annealing, also referred to as adiabatic quantum computation (10-13), as well as new protocols for generating quantum cluster states (14).

The $LiHo_xY_{1-x}F_4$ family of insulating magnetic salts provides a physical manifestation of the simplest quantum mechanical spin model, the Ising model in transverse field (15). Pure $LiHoF_4$ (16, 17) is a ferromagnet with Curie temperature, $T_C = 1.53$ K. External magnetic fields can produce the longitudinal and transverse fields in the model, chemical substitution of $Ho^{3+}$ ions by the non-magnetic species $Y^{3+}$ provides quenched disorder, and the anisotropy of the dipolar coupling produces random internal transverse fields (18-21) as well as competing ferromagnetic and antiferromagnetic interactions. The combination of site dilution and external fields yields a wide variety of collective magnetic states, ranging from random field ferromagnet at x = 0.44 (22) to quantum spin glass at x = 0.167 (23). We focus here on the dilute limit of x = 0.045, for which there have been seemingly contradictory findings concerning the ground state.

The primary diagnostic of the ground state has been the ac magnetic susceptibility, whose imaginary part $\chi''(f)$ is the quotient of the long-wavelength magnetic fluctuation spectrum, $S(f)$, and the Bose factor , $(n(hf)+1)=1/(1-\exp{-hf/kT})$, where $h$ and $k$ are Planck's and Boltzmann's constants, respectively. For our experiments, $hf \ll kT$ and hence $\chi''(f) = hf/kT \, S(f)$. The frequency at which the imaginary part peaks indicates the characteristic relaxation rate of the system, which for spin dynamics dominated by thermal activation over energy barriers will vary in accord with the Arrhenius law, $f = f_0 \exp(-E_A/kT)$ (15, 24, 25), precisely what we see for temperatures 0.15 K < T < 1 K. Below T ~ 0.15 K, deviations from Arrhenius behavior emerge (15, 24, 25); however, the nature of the deviations and their interpretation has been contested (26). One class of experiments found a low-frequency narrowing of the spectrum with decreasing T (24, 27, 28), accompanied by the magnetic equivalent of optical hole burning in the non-linear response, where effectively isolated, mesoscopic clusters of spins can be addressed and manipulated using a pump/probe technique (24). A magnetic field applied transverse to the Ising axis introduces quantum fluctuations, and can influence the relaxation pathways of the coherent clusters (28). Moreover, muon spin relaxation (µSR) studies have shown that the persistent spin fluctuation rate remains constant down to T = 0.02 K, consistent with a spin-liquid ground state (29). By contrast, a second class of magnetic susceptibility studies found that LiHo$_{0.045}$Y$_{0.955}$F$_4$ behaved as a paramagnet approaching a spin-glass transition which extrapolation suggests to occur at T$_g$ ~ 0.04 K, with a magnetic fluctuation spectrum that broadened symmetrically as the temperature was lowered (25). In this picture, the characteristic dissipative response moves more quickly to low frequency as the system as a whole freezes.

The most significant distinction between the two classes of susceptibility experiments is the heat sinking of the sample to the cryostat. For the measurements yielding a spin liquid, a single crystal measuring (5 x 5 x 10) mm$^3$ was heat-sunk by sapphire rods pressed against the sample on either end of the long axis (24); in the spin-glass case a (0.57 x 0.77 x 7.7) mm$^3$ sample was glued to a sapphire rod running along its length (25). The sapphire rods are then thermally anchored to the mixing chamber of the dilution refrigerator, coupling them to the environmental heat bath. If the thermal boundary conditions of the sample change appreciably, then the internal state of the system also may be expected to change. Just as the application of a transverse magnetic field affects the spin cluster dynamics and their coupling to the external world in this system (28), thermal boundary conditions can enhance or destroy isolated spin degrees of freedom, tune the system between classical and quantum mechanical limits (30), and alter the relative energies of competing ground states.

**Results and Discussion**

To test whether such tuning can be realized experimentally, we measured the linear and non-linear ac magnetic susceptibility of the same LiHo$_{0.045}$Y$_{0.955}$F$_4$ crystal studied in Ref. (24), but now in two different thermodynamic limits (Fig. 1A). For the first configuration (Fig. 1A left), the sapphire heat-sinking rods are in direct, spring-loaded contact with the sample on each end, and are mechanically connected (in vacuo) to the cryostat cold finger on the other. In this "strongly coupled" geometry, the ends of the sample define lattice isotherms. In the second configuration, the sapphire rods are mechanically separated from the sample with a 4 mm average vacuum gap by using Teflon spacers at the rear of the sapphire rods (Fig. 1A right). In this "weakly coupled" geometry, the dominant thermal link between the sample and the cryostat is through the Hysol (epoxy resin) body of the susceptometer coil form. The thermal conductivity of the sapphire at $T = 0.1$ K allows an energy flow of approximately 10 nW/mK between the sample and cryostat, to be compared with that of the disconnected geometry's 0.3 nW/mK, and black-body heat flow from the vacuum can at $T = 4.2$ K of $1 \times 10^{-8}$ nW/mK, much less than the heat flow through the Hysol coil form in the weakly coupled geometry. The helium dilution refrigerator itself has approximately 300 µW of cooling power at 0.1 K. The ac probe fields applied in the mutual inductance coils were 40 mOe or below so as to be in the linear response region. The Earth's static magnetic field was not shielded.

As the crystal is cooled from high temperature, the ways in which the spin configurations evolve differs for the two connections to the thermal bath. In the strongly coupled case, heat is rapidly extracted from the nuclear spin and phonon bath in contact with the Ho spins, meaning that energy-lowering spin flips are available throughout the sample to thermally surmount barriers until the lowest T. This would lead to spins gradually forming a connected, disordered network of frozen spins, as illustrated in Fig. 1B, if this were indeed the true ground state. At the same time, dissipative interactions between the quantum spin system and the thermal bath imply that the motions of the principal electronic and nuclear degrees of freedom associated with Ho ions do not conserve energy. Pure quantum eigenstates of the combined electronuclear spin

Hamiltonian of LiHo$_x$Y$_{1-x}$F$_4$, would then have shorter coherence times, eliminating their influence on observables such as the magnetic susceptibility(2, 31). By contrast, in the weakly coupled case, the magnetic energy is effectively conserved because there is little heat transport away from the sample and the spins evolve according to the quantum spin Hamiltonian (31). A system comprised of an ensemble of these clusters of spins superposed upon an incoherent bath of spins, also sketched in Fig. 1B, would give rise to stronger quantum characteristics such as entanglement and zero point fluctuations at low T, as might be manifested by the magnetic susceptibility (18), persistent μSR spin fluctuation rates (29), and Fano resonances in non-linear bleaching experiments (28).

In the strongly coupled configuration, we indeed find that the system behaves as a classical glass (Fig. 1D left). The peak frequency in $\chi''$ follows an Arrhenius law down to at least 125 mK with activation energy $E_A$ = 1.54 ± 0.03 K, in agreement with the result $E_A$ = 1.57 K reported in the literature (25) and sufficiently above any possible glass transition that a Vogel-Fulcher-type $T \rightarrow T - T_0$ correction (32) to Arrhenius behavior would not be visible. In contrast, for weak coupling to the bath we observe pronounced deviations from Arrhenius behavior at low T in good agreement with the results of Ref. (18).

An important question is whether the disparate results correspond to genuinely different magnetic states of the LiHo$_x$Y$_{1-x}$F$_4$ crystal or simply follow because the effective sample temperatures are different, perhaps because of inadequate cooling in the weakly coupled case. Both the linear and non-linear response unambiguously support the former hypothesis. In particular, if we were dealing with a simple cooling problem as suggested by pure Ising simulations (without internal transverse fields generated by off-diagonal terms of dipolar interactions) (33), then the peak frequency for $\chi''$ could be used as a proxy for the sample temperature and spectra with the same peak frequency should overlay perfectly. Fig. 1D (center) shows that this is not the case, with weak coupling resulting in a spectrum with a suppressed low-frequency tail, at odds with the broader tail seen for strong coupling, which is consistent with data in Ref. (25).

We have checked as well for time-dependence of the approach to equilibrium in both cases. Fig. 2, which illustrates how the 10 Hz response approaches its 90 mK equilibrium value after a quench from 9 K, reveals that for both weak and strong bath couplings, no time dependence is discernible after $10^4$ seconds. Equally significant is that there are multiple crossings of the $\chi''$ data for the two thermal couplings, meaning that the isolated response is not simply a time-extended version of the connected response. In particular, at long times, there is a crossing to apparently different equilibrium values of the imaginary part of the magnetic susceptibility.

We note that the relaxation times shown in Fig. 2 are 2 to 4 orders of magnitude longer than the 1 to 100 second characteristic time scales of the collective spin modes after the relaxation is complete (cf. the 10 mHz-1 Hz peak frequencies for the different coupling cases at T = 90 mK in Fig. 1D). Studies of single-$Ho^{3+}$ physics in highly dilute $LiHo_xY_{1-x}F_4$ have shown that spin-phonon coupling and the relaxation rates of the phonons impact the single-spin behavior (34), and a similar effect is likely driving the collective multispin modes studied here. The time scales arise from the presence of a "phonon bottleneck" (35, 36), where the relatively slow phonon dynamics act as a rate-limiting step for the equilibration of the spin dynamics. The thermal coupling impacts the rate at which the phonons can be used to extract heat from the spin system and hence the details of quantum-mechanical level crossings associated with the electronuclear spin Hamiltonian are of greater importance for weaker coupling; a similar effect was observed as a function of thermal coupling for the S=1/2 molecular magnet $V_{15}$ (37).

Fig. 1C illustrates the underlying phenomena. First, consider three subsystem (e.g. a cluster at right in Fig. 1B) states, 1, 2 and 3 which, for a bath characterized by the temperature-dependent parameter $\alpha(T)$, exchange places in their ranking according to the classical potential energy as T is lowered to yield $\alpha(T) = \alpha 1, \alpha 2, \alpha 3$. This potential energy is defined by other degrees of freedom which may be undergoing thermal fluctuations that are slow on the scale of the tunneling times for the three levels being considered (12). We now introduce quantum mixing, for example via a transverse field generated either internally or externally. The potential energy curves are drawn so that the minima for states 1 and 2 are closer in phase space to each other than to the minimum for state 3, meaning that within a barrier tunneling picture (38) there will be more mixing of states 1 and 2 than of either of these states with state 3. In such a circumstance, quantum mixing actually can lead to a condition where at $\alpha 3$ the subsystem state will be predominantly a mixture of states 1 and 2 (blue circle) even while classically the system will clearly be in state 3 (red circle). Quantum mixing will yield such a result depending on whether the cooling rate is faster than the decoherence rate produced by the environment not incorporated in the Hamiltonian including the mixing terms. These decoherence times are fixed by the rate at which phonons can be supplied to the bath, which is in turn controlled by the heat sinking.

Another diagnostic of the magnetic state is the probe amplitude-dependent response, where we observe how the system departs from the linear, small signal regime. The right hand side of Fig. 1D demonstrates a profound difference between the nonlinear susceptibilities for the two couplings: for the adiabatic limit, we observe an s-like Brillouin term superposed on a linear background, while for strong coupling to the bath, the linear regime persists to higher drive field, after which the response grows more dramatically to cross that for weak coupling to the bath.

Previous work has shown that transverse fields $H_t$ applied to $LiHo_xY_{1-x}F_4$ induce both quantum fluctuations and static random fields. Given that the random field amplitudes and quantum tunneling rates scale as $H_t$ and $H_t^2$ respectively (21, 22), random field effects, including pinning and associated slowdown of spin fluctuations, should dominate at small $H_t$. As $H_t$ is increased beyond the crossover field where tunneling rates and pinning energies are in balance, quantum speedup should become visible. A crossover between regimes dominated by $H_t$-induced longitudinal random fields and $H_t$-induced tunneling was in fact observed in the more concentrated random ferromagnet, $LiHo_{0.46}Y_{0.56}F_4$ (22, 39). For weak coupling, Fig. 3A displays precisely the anticipated effects even in the x = 0.045 sample, namely a softening of the spectrum as we raise $H_t$ from zero to 100 Oe, and then a hardening upon further increasing $H_t$ to 2000 Oe. Consistent with classical pinning, the initial softening effect is visible only when cooling in field from high temperature, whereas the high transverse field (2 kOe) hardening is history-independent and therefore consistent with $H_t$-induced quantum speedup.

Fig. 3B-C, showing the behavior of the spin liquid after preparation via different trajectories through $H_t$-T space, supports the random field-quantum crossover description for weak heat sinking. In addition to the quantum annealing protocol (i), we also followed the mixed $H_t$/thermal schedules (ii) where the transverse field $H_t^0$ is applied before cooling, but not removed at lowest T as in (i), and (iii) where the transverse field is only applied after cooling to the lowest T. Despite the data being collected at identical (T, $H_t^0$), the spectra do not coincide below $H_t^0 \sim 500$ Oe. The classical pathway (iii) becomes unstable over the course of a day for $H_t^0 \sim 500$ Oe, and the two protocols merge over repeated measurements of the spectrum. Above this field scale, the transverse field is large enough to produce the quantum fluctuations required to overcome pinning of spin configurations in the search for equilibrium. For even larger transverse fields ($H_t > 3.5$ kOe), the physics of quantum level crossings for single ions and pairs of ions, determined both by purely electronic and nuclear hyperfine interactions, emerges and can be probed directly(28, 40).

An important result is that tuning the system state with $H_t^0$ requires thermal cycling above T = 9 K before changing the transverse field. Warming above the Curie temperature of the pure compound, $T_C = 1.53$ K, or even to T = 4 K, is not sufficient to reset the system upon cooling. Rather, the low temperature state only responds to a change in $H_t^0$ if the spins are thermally excited above the 9.4 K splitting between the Ising ground state doublet and the first excited state singlet (15). This points once again to the fundamental quantum nature of the ground state when isolated from the incoherent thermal bath.

We explore in Fig. 4 the spectral characteristics of the weakly coupled spins after quantum annealing, probing both the linear and non-linear response. There appear to be a continuous set of low temperature states that the system can access depending on the strength of the cooling field. For $H_t^0 < 0.1$ kOe, we see that $\chi''$ behaves as in Fig. 3A, where the cooling field was maintained to achieve the final state. In particular, there is a softening with increasing transverse field, corresponding to the random fields present on the cool-down which pin the spin configuration even after the transverse field is shut off at low T. On the other hand, the final state is much less dependent on $H_t^0 > 0.1$ kOe, consistent with the capability of quantum annealing to produce approximately the same final state independent of the transverse field during cooling; the discrepancies found at low frequencies may be linked to small differences in precisely how the random field regime for $H_t < 0.1$ kOe was traversed.

We have shown that for weak thermal coupling, LiHo$_{0.045}$Y$_{0.955}$F$_4$ approaches a different quasi-equilibrium state than for strong thermal coupling. In particular, the former state seems to display more apparently quantum mechanical traits, most notably a spectrum with a higher characteristic frequency than predicted from thermal activation, a nearly dissipationless response at low frequencies, and more acute sensitivity to externally applied transverse fields (Fig. 4B). As in optical spectroscopy, pump/probe studies of LiHo$_{0.045}$Y$_{0.955}$F$_4$ can be used to determine the extent to which the response spectrum is due to an inhomogeneous mixture of different oscillators with different resonance frequencies or to relaxation of coupled degrees of freedom. For strong thermal coupling, we simply observe the earmarks of heating when applying a large (~0.3 Oe) pump field. By contrast, for weak thermal coupling we observe a Fano resonance centered at the pump frequency (Fig. 4A inset), corresponding to a discrete oscillator coupled to a continuum:

$$\Delta\chi'' \propto \frac{\left(\frac{Q\Gamma}{2} + \Delta f\right)^2}{\Delta f^2 + \left(\frac{\Gamma}{2}\right)^2} \qquad (1)$$

where $\Delta f$ is the distance from the center point of the resonance, $\Gamma$ is the characteristic width of the resonance and the Fano parameter $Q$ parameterizes the asymmetry of the resonance. The Fano effect is a quantum interference effect, first described to explain the absorption spectra of gas molecules (41) and later observed in a wide variety of quantum two-level systems ranging from semiconductor quantum wells (42) to the BEC-BCS crossover in cold atom experiments (43, 44). In LiHo$_{0.045}$Y$_{0.955}$F$_4$, the presence of this resonance for a weakly-linked thermal bath, and absence for a strongly-linked bath, again confirm the need for a quantum mechanical description of the ground and low-lying excited states for weak linkage to the bath. We plot in Fig. 4C the results of fits of Eq. 1 to our data. They reveal that while the coupling constant Q is essentially independent of the transverse field applied while cooling, the amplitude of the resulting Fano resonance, and hence the number of oscillators coupled to the bath, can be tuned continuously by $H_t^0$. The latter undergoes a type of phase transition at a critical $H_{tc}^0 \sim 0.1$ kOe identified in the linear response experiments. It varies as $C + (H_{tc}^0 - H_t^0)^\alpha$ below 0.1 kOe with $\alpha = 1.5\pm0.1$, and matches the constant C in remarkable field-independent fashion above 0.1 kOe. This means that some local oscillators, most likely associated with the motion of spins near pinning sites, survive the larger quantum fluctuations brought about by the larger transverse field.

Our results have both fundamental and practical implications. On the fundamental side, they mean that even for macroscopic solid-state systems, thermal coupling has effects similar to those of stopping gases that destroy quantum effects in atomic physics experiments via collisions. Weakening the thermal coupling converts a classical glass into a stable system with the earmarks of a quantum spin liquid. The practical implication is that the performance of quantum annealing machines, of which our apparatus is one example (used here to solve the specific problem of finding the most probable spin states for a crystal of LiHo$_{0.045}$Y$_{0.955}$F$_4$), can depend strongly on the coupling to their heat baths.

**Acknowledgments**

The work at The University of Chicago was supported by DOE Basic Energy Sciences Grant No. DE-FG02-99ER45789 and in London by the UK Engineering and Physical Sciences Research Council. GA is grateful to N. Chancellor for discussions.

**Figure Legends**

**Fig. 1**: **Effects of thermal boundary conditions on LiHo$_{0.045}$Y$_{0.955}$F$_4$.** **A** Schematic of experimental arrangement. The sample sits at the center of an ac susceptometer coil set. Sapphire rods connected via copper wires to the cryostat cold finger provide thermal contact to the sample. (left) Rods in direct contact with sample (right) Teflon spacers at the backs of the rods impose a 4 mm vacuum gap between rods and sample, making the epoxy resin coil assembly the dominant route for thermalization. **B** Cartoon of spin configurations inside the LiHo$_{0.045}$Y$_{0.955}$F$_4$ crystal for the two experimental configurations in A. (left) In the presence of strong thermal coupling to a heat reservoir, the system forms a glassy network dominated by thermal fluctuations. (right) In the absence of a strong connection to the reservoir, isolated spin clusters with discrete quantum transitions coupled weakly to the continuum of excitations of other clusters are an important feature of the low temperature response. **C** At top is a schematic showing the thermal evolution (associated with the mean field from other clusters) of the potential energy for different states of a typical spin cluster as shown in B. Bottom frame shows how quantum mixing of states alters the outcome of the cooling process, with a quantum mixture of eigenstates 1 and 2 rather than the lowest energy state 3 emerging for cooling rapid compared to decoherence rates. **D** Measured effects of changing thermal coupling conditions. (Left) Peak frequency of imaginary susceptibility as a function of inverse temperature for the two configurations drawn in A, compared with values published in Refs (24) and (25). (Center) Lineshapes of the imaginary susceptibility under different thermal conditions for the same peak frequency. Measurements in the weakly coupled thermal configuration show low-frequency spectral narrowing not observed in the connected configuration. (Right) Magnetization curves at fixed frequency for the two thermal configurations, showing markedly different functional forms and field scales for the onsets of nonlinearity and saturation.

**Fig. 2: Time evolution after thermal quenching with different couplings to the bath.** Real and imaginary susceptibilities for the two thermal configurations in Fig. 1A at T = 0.09 K and f = 10 Hz as a function of time t following a thermal quench from 9 K. Time t=0 marks the time at which χ' peaks. The trajectories are not only different, but appear to reach equilibria characterized by different values of the susceptibility.

**Fig. 3**: **Random-field pinning and quantum speedup with transverse field.** **A** Imaginary susceptibility of LiHo$_{0.045}$Y$_{0.955}$F$_4$ with weak coupling to the bath after cooling from 9 to 0.09 K in a transverse field $H_t^0$. The spectrum moves non-monotonically with $H_t^0$, first slowing from enhanced random field pinning and then moving to higher frequency via transverse field-induced tunneling. Dashed line at 3.1 Hz indicates constant-frequency cut used in panel C. **B** Trajectories through $H_t$-T space. **C** Constant-frequency response of the imaginary part of the susceptibility following the three different cooling trajectories over a range of transverse fields. In the quantum annealing trajectory (i), as in (ii) where the quantum fluctuation rate remains constant approaching the final state, $H_t^0 \sim 0.1$ kOe demarcates different response regimes. The classical annealing trajectory (iii) shows a weak quantum speedup at low fields followed by an instability at fields above 0.5 kOe that yields a tunneling transition to branch (ii) with a timescale ~ 1 day.

**Fig. 4**: **Spectral character of the linear and nonlinear response. A** Imaginary part of the susceptibility from 1 Hz to 1 kHz after quantum annealing in a series of transverse fields with the crystal weakly coupled to the bath. The curve in black is the dissipative response of the sample in the strongly coupled limit with $H_t^0 = 0$. An *in situ* GaAs Hall magnetometer was used to directly measure the applied transverse field. Inset: Non-linear pump-probe spectroscopy for the same quantum annealing trajectories as in the main panel, with a 19.95 Hz/0.3 Oe pump field. The Fano resonance fits (smooth curves, Eq. 1 in the text) indicates interference between excitations of discrete spin cluster states and those in the bath; dashed lines show predicted behavior for the resonances for Δf less than the experimental resolution. **B** Constant-frequency (f=3.1 Hz) cuts of the imaginary susceptibility as a function of the cooling field for the two couplings to the bath. **C** The Fano coupling parameter Q (see Eq. 1 in the text) is essentially independent of cooling field, but the amplitude of the resonance, and hence the number of discrete oscillators coupled to the bath, decreases until the demarcation field $H_{tc}^0 \sim 0.1$ kOe identified in Fig. 3C, and then stabilizes at a finite value.

# Figures

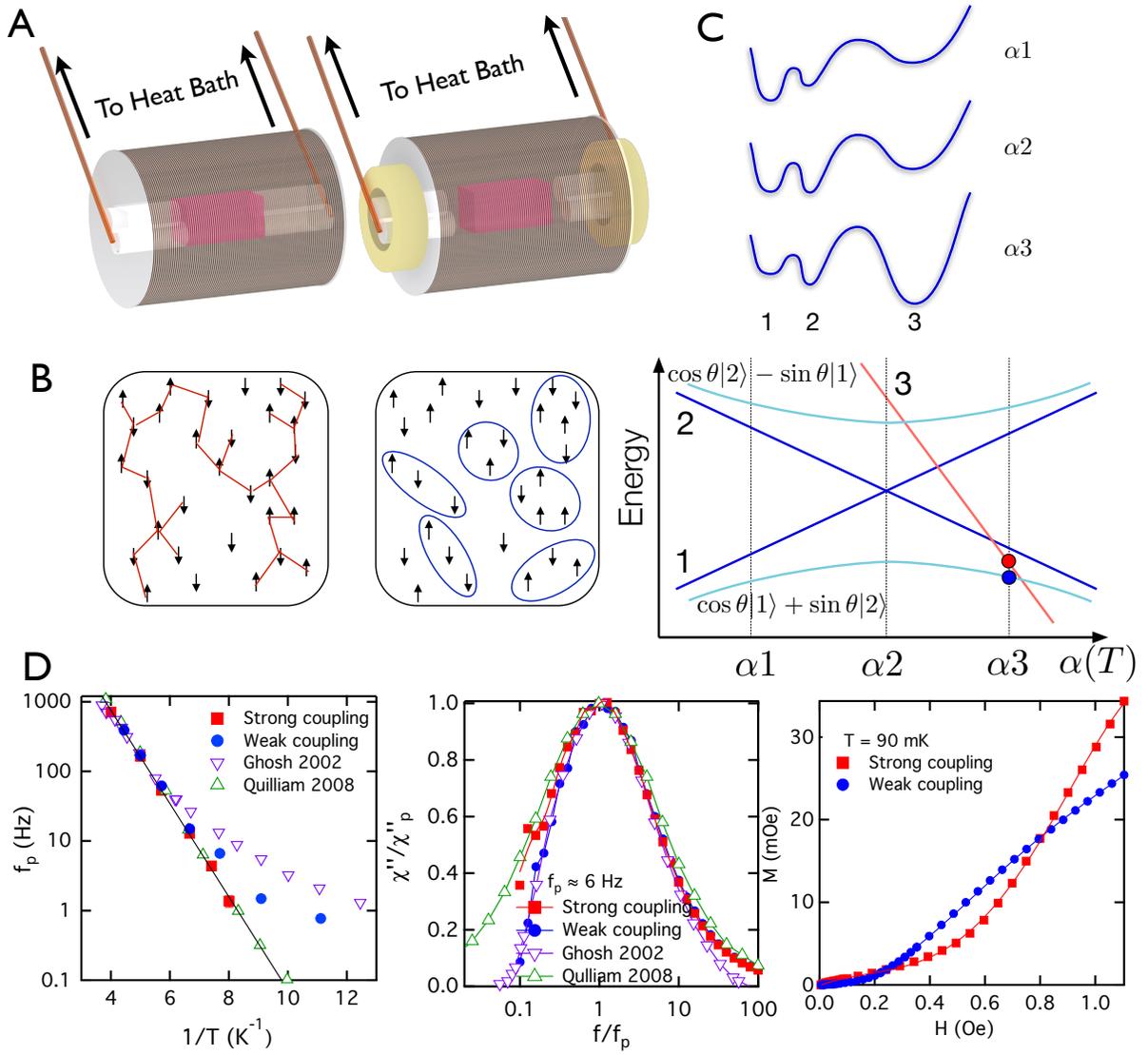

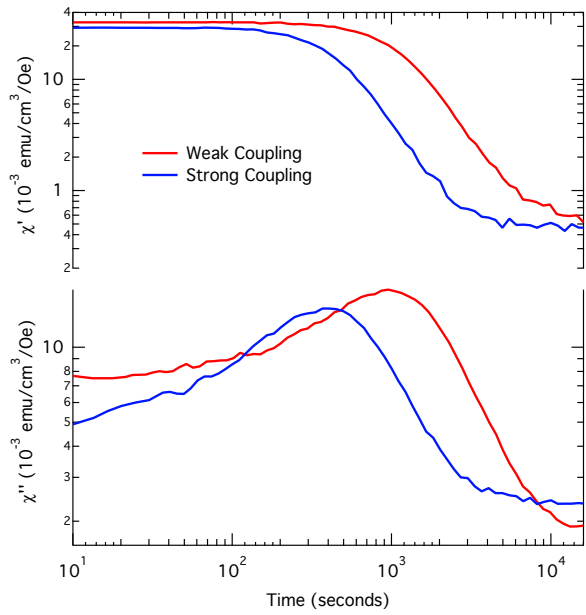

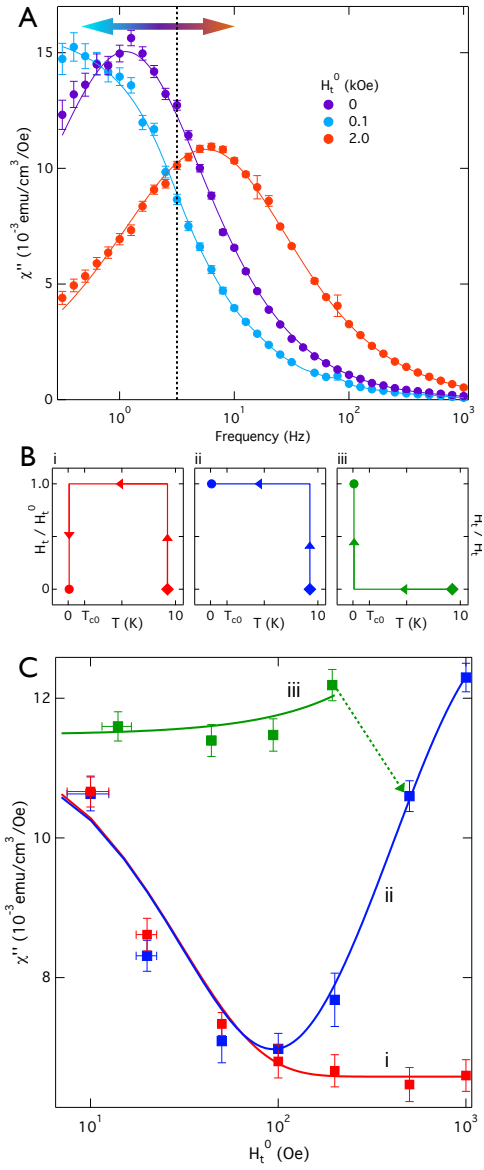

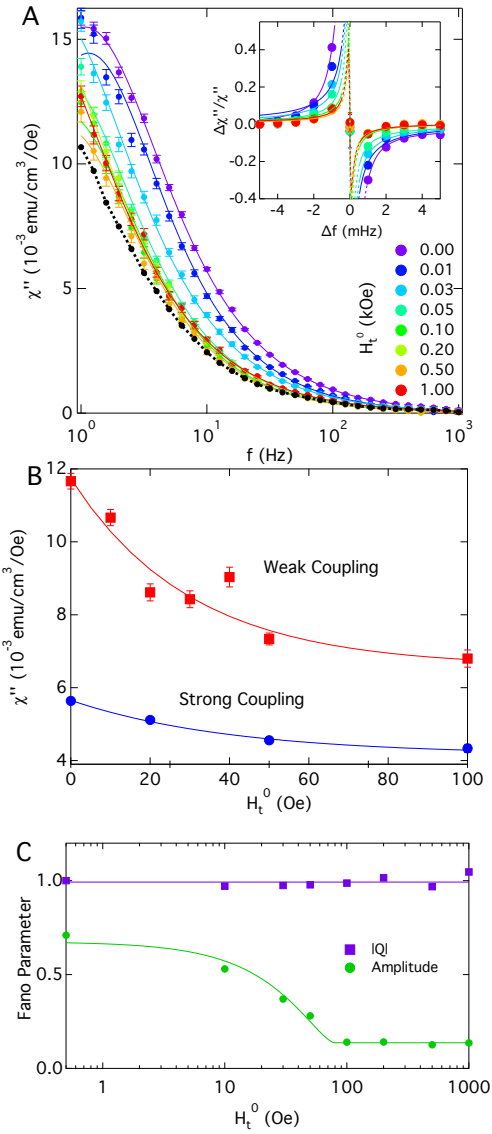